\documentclass[twocolumn,showkeys,aps,prb,showpacs]{revtex4-1}
\usepackage{graphicx}
\usepackage{amsmath}
\usepackage[CJKbookmarks,dvipdfm,colorlinks,linkcolor=blue,citecolor=blue]{hyperref}

\begin{document}

\title{Predicted Janus monolayer ZrSSe with enhanced n-type thermoelectric properties compared with monolayer  $\mathrm{ZrS_2}$}

\author{San-Dong Guo}
\affiliation{School of Physics, China University of Mining and
Technology, Xuzhou 221116, Jiangsu, China}

\begin{abstract}
In analogy to  transition-metal dichalcogenide (TMD) monolayers,  which have wide applications in photoelectricity, piezoelectricity and thermoelectricity,
 Janus MoSSe monolayer has been successfully synthesized  by substituting  the top Se atomic layer in  $\mathrm{MoSe_2}$ by S atoms. In this work,
Janus monolayer ZrSSe is proposed by ab initio calculations.  For the electron part, the generalized gradient approximation (GGA) plus spin-orbit coupling (SOC)  is used as exchange-correlation potential, while  GGA for  lattice part. Calculated results show  that the ZrSSe monolayer is dynamically and
mechanically stable, which exhibits mechanical flexibility due to small Young's modulus. It is found that ZrSSe monolayer is an indirect-gap semiconductors
with band gap of 0.60 eV. The electronic and  phonon transports of ZrSSe monolayer are  investigated by  semiclassical Boltzmann transport theory.
In n-type doping, the $ZT_e$ between ZrSSe and $\mathrm{ZrS_2}$ monolayers is almost the same due to similar outlines of conduction bands. The p-type $ZT_e$ of ZrSSe monolayer is lower than that of $\mathrm{ZrS_2}$ monolayer, which is due to larger  spin-orbit splitting for ZrSSe than $\mathrm{ZrS_2}$ monolayer.
The room-temperature sheet thermal conductance  is 33.6 $\mathrm{W K^{-1}}$ for ZrSSe monolayer,  which is lower than  47.8 $\mathrm{W K^{-1}}$ of $\mathrm{ZrS_2}$ monolayer. Compared to  $\mathrm{ZrS_2}$ monolayer, the low sheet thermal conductance of ZrSSe monolayer is mainly due to small group  velocities and short phonon lifetimes of ZA mode. Considering their $ZT_e$ and  lattice thermal conductivities, the ZrSSe monolayer may have better n-type thermoelectric performance than $\mathrm{ZrS_2}$ monolayer. These results can  stimulate further experimental works to synthesize  ZrSSe monolayer.

\end{abstract}

\pacs{72.15.Jf, 71.20.-b, 71.70.Ej, 79.10.-n ~~~~~~~~~~~~~~~~~~~~~~~~~~~~~~~~~~~Email:sandongyuwang@163.com}

\maketitle

\section{Introduction}
 Thermoelectric generators with no moving parts are silent, reliable and scalable,  having potential  applications in energy-related issues
\cite{q0,q1,q4,q6}. The performance of a
thermoelectric device is  measured by the thermoelectric
material's dimensionless figure of merit
\begin{equation}\label{aa}
    ZT=S^2\sigma T/(\kappa_e+\kappa_L)
\end{equation}
in which S, $\sigma$, T, $\kappa_e$ and $\kappa_L$ are the Seebeck coefficient, electrical conductivity, working temperature,  electronic and lattice thermal conductivities, respectively. Identifying materials with high thermoelectric efficiency is challenging  by searching for  a high power factor ($S^2\sigma$) and/or a low thermal conductivity ($\kappa=\kappa_e+\kappa_L$), which is  due to the conflicting combination of  these transport coefficients.
For metals or degenerate
semiconductor, the  S and  $\sigma$ are given by\cite{new1}:
\begin{equation}\label{aa}
    S=\frac{8\pi^2K_B^2}{3eh^2}m^*T(\frac{\pi}{3n})^{2/3}
\end{equation}
\begin{equation}\label{aa}
    \sigma=ne\mu
\end{equation}
where $n$, $m^*$ and $\mu$ is the carrier concentration, the effective mass of
the carrier and carrier mobility, respectively. It is clearly seen that  the  S and  $\sigma$ are oppositely  proportional to $n$.

Firstly proposed by Hicks and Dresselhaus in 1993\cite{q2,q3}, the low-dimensional
systems or nanostructures could have much higher $ZT$ values than their bulk counterparts.
A large variety of two-dimensional (2D) monolayers beyond graphene have been predicted in theory, or  synthesized experimentally,
such as  TMD\cite{q7},  Janus TMD\cite{q7-1},  group IV-VI\cite{q8}, group-VA\cite{q9,q10}, group-IV\cite{q11}  monolayers and so on . The  heat transports of these 2D monolayers  have been widely investigated theoretically or experimentally.
\begin{figure}
  \includegraphics[width=6cm]{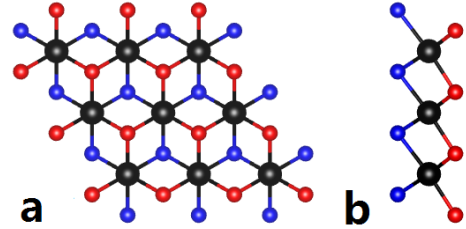}
  \caption{(Color online) (a) Top and (b) side view of the crystal structure
of ZrSSe. The large black balls represent Zr atoms, and the small red and  blue balls for S and Se atoms.}\label{st}
\end{figure}
  The thermoelectric properties of $\mathrm{MX_2}$ (M=Mo or W; X=S or Se),  orthorhombic group IV-VI  $\mathrm{AB}$ (A=Ge or Sn; B=S or Se), As, Sb, Bi, $\mathrm{SnSe_2}$, $\mathrm{TiS_2}$  and ATeI (A=Sb or Bi) monolayers  have been  systematically investigated  theoretically\cite{q12,q15,q16,q16-1,q21,q22,q14,q14-1,q14-2,q14-3}.
  For monolayer $\mathrm{MoS_2}$, the S  of 30 mV/K  has been reported experimentlly\cite{q13}.
The  phonon transport properties of  group-IV, ZnO, GaN and SbAs  monolayers have been systematically investigated  from ab initio calculations\cite{q23,q24,q25,q26}. Strain dependent phonon transports of 2D Penta-Structures, antimonene, silicene, germanene and stanene have been studied by solving the phonon Boltzmann transport equation\cite{q27-1,q27-2,q27-3}.

\begin{figure}
  \includegraphics[width=8cm]{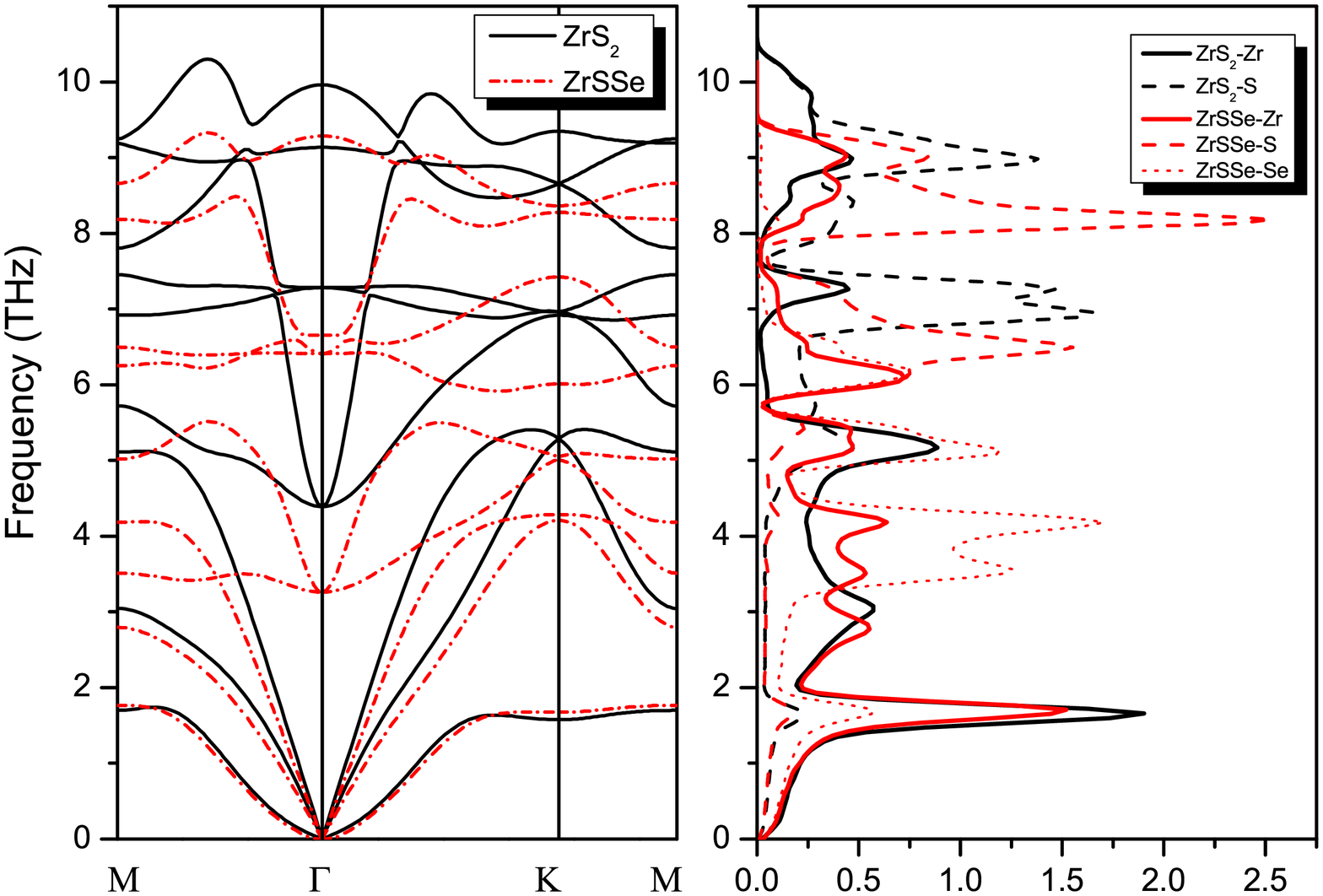}
  \caption{Phonon dispersion curves of $\mathrm{ZrS_2}$ and ZrSSe monolayers with the
corresponding atom partial DOS using GGA-PBE.}\label{phband}
\end{figure}

Recently, the Janus monolayer MoSSe, breaking   inversion and mirror symmetry , has been confirmed  by means of scanning transmission
electron microscopy and energy-dependent X-ray photoelectron
spectroscopy\cite{q7-1}. Based on ab initio calculations, the large in-plane and vertical piezoelectricity in
Janus TMD MXY (M =Mo or W, X/Y = S, Se or Te) has been predicted, which  reveals the potential for utilizing piezoelectric 2D materials\cite{n1}.
Electronic and optical properties of pristine and  vertical and lateral heterostructures of Janus MoSSe and WSSe monolayers have been reported on the basis of electron-electron self-energy corrections\cite{n2}. The  Janus  monolayer related with 2H structure TMD  has been widely studied, and it is amusing to explore  1T Janus TMD  monolayer.  The representative 1T TMD is  $\mathrm{ZrS_2}$, which has
been successfully synthesized experimentally\cite{n3}. The $\kappa_l$ of  $\mathrm{ZrS_2}$ monolayer is predicted to be
much lower than those of $\mathrm{MX_2}$ (M = Mo or W; X = S or Se) monolayers\cite{q8}. Strain-induced enhancement of thermoelectric
performance in   $\mathrm{ZrS_2}$ monolayer has been predicted,  based on first-principles calculations combined with the
Boltzmann transport theory\cite{n4}.

To further improve thermoelectric performance of $\mathrm{ZrS_2}$ monolayer, Janus monolayer ZrSSe is proposed, which can be constructed  by substituting  the top S atomic layer in  $\mathrm{ZrS_2}$ by Se atoms. It is found that the ZrSSe monolayer  is dynamically and
mechanically stable, exhibiting  mechanical flexibility. An indirect-gap semiconductor is observed in ZrSSe monolayer, with band gap of 0.60 eV using GGA+SOC.
Calculated results show that, in n-type doping, the $ZT_e$ between ZrSSe and $\mathrm{ZrS_2}$ monolayers is almost the same due to similar outlines of conduction bands. The room-temperature sheet thermal conductance  (33.6 $\mathrm{W K^{-1}}$) of ZrSSe monolayer,   is lower than that (47.8 $\mathrm{W K^{-1}}$) of $\mathrm{ZrS_2}$ monolayers, which is mainly due to smaller  group  velocities and shorter phonon lifetimes of ZA mode.
Considering their $ZT_e$ and $\kappa_L$, the ZrSSe monolayer may have better n-type thermoelectric performance than $\mathrm{ZrS_2}$ monolayer.

The rest of the paper is organized as follows. In the next
section, the  computational details about  electronic structures, electron and phonon transports are given. In the third section, we shall present elastic properties, electronic structures, electron and  phonon transports of  ZrSSe monolayer. Finally, we shall give our discussions and conclusions in the fourth section.
\begin{figure}
  \includegraphics[width=8cm]{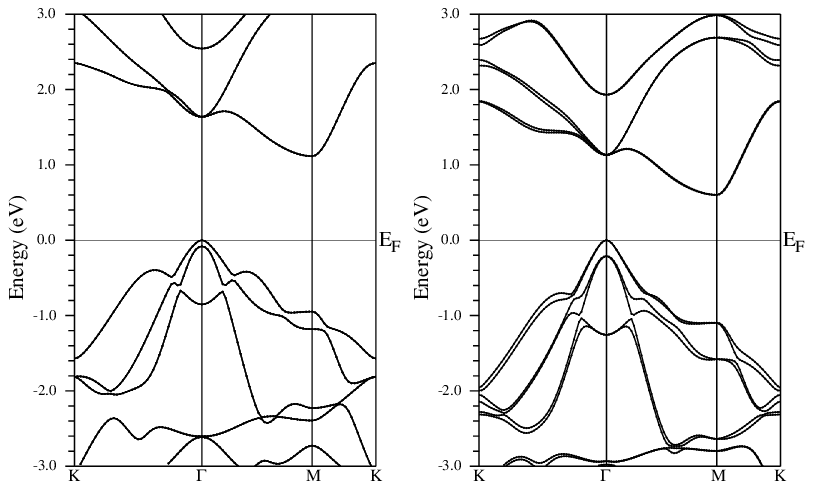}
  \caption{The energy band structures of  $\mathrm{ZrS_2}$ (Left) and ZrSSe (Right) monolayers using GGA+SOC.}\label{band}
\end{figure}

\begin{table}[!htb]
\centering \caption{For $\mathrm{ZrS_2}$ and ZrSSe monolayers, the  lattice constants $a$ ($\mathrm{{\AA}}$); the Zr-S and Zr-Se bond lengths $d$ ($\mathrm{{\AA}}$); the energy band gap $G$ (eV);  the value  of spin-orbit splitting at $\Gamma$ point $\Delta$ (eV).}\label{tab}
  \begin{tabular*}{0.48\textwidth}{@{\extracolsep{\fill}}ccccccc}
  \hline\hline
Name& $a$ &  $d_{Zr-S}$& $d_{Zr-Se}$&$G$& $\Delta$\\\hline\hline
$\mathrm{ZrS_2}$&3.68&2.58&-&1.12& 0.09\\\hline
ZrSSe&3.74&2.57&2.71&0.60&0.21\\\hline\hline
\end{tabular*}
\end{table}
\begin{table*}
\centering \caption{For $\mathrm{ZrS_2}$, ZrSSe,  $\mathrm{MoS_2}$ and MoSSe monolayers, the elastic constants $C_{ij}$, shear modulus
$G^{2D}$,  Young's modulus $Y^{2D}$ in $\mathrm{Nm^{-1}}$, and Poisson's ratio $\nu$
dimensionless. }\label{tab1}
  \begin{tabular*}{0.96\textwidth}{@{\extracolsep{\fill}}ccccccc}
  \hline\hline
Name& $C_{11}=C_{22}$ &  $C_{12}$& $C_{66}=G^{2D}$&$Y_{[10]}^{2D}=Y_{[01]}^{2D}$& $\nu_{[10]}=\nu_{[01]}$\\\hline\hline
$\mathrm{ZrS_2}$&75.96&16.23&29.82&72.49& 0.21\\\hline
ZrSSe&68.84&14.72&27.06&65.69&0.21\\\hline\hline
$\mathrm{MoS_2}$&138.5&31.7&53.4&131.2& 0.23\\\hline
MoSSe&126.8&27.4&49.7&120.9&0.22\\\hline\hline
\end{tabular*}
\end{table*}

\section{Computational detail}
The electronic structures of  $\mathrm{ZrS_2}$ and ZrSSe  monolayers are calculated
by a full-potential linearized augmented-plane-waves method
within the density functional theory (DFT)\cite{1}, as implemented in the  WIEN2k code \cite{2}.
The GGA of Perdew, Burke and  Ernzerhof  (GGA-PBE)\cite{pbe} is used as  the exchange-correlation functional, and  the free  atomic position parameters  are optimized  with a force standard of 2 mRy/a.u..
The SOC is included self-consistently \cite{10,11,12,so}. The convergence results are determined
by using  4000 k-points in the
first Brillouin zone (BZ) for the self-consistent calculation, making harmonic expansion up to $\mathrm{l_{max} =10}$ in each of the atomic spheres, and setting $\mathrm{R_{mt}*k_{max} = 8}$ for the plane-wave cut-off.  From calculated energy band
structures, the electronic  transport coefficients are performed  through solving Boltzmann
transport equations within the constant
scattering time approximation (CSTA),  as implemented in
BoltzTrap code\cite{b}. To obtain accurate transport coefficients,  the parameter LPFAC  is set as 20, and  at least 2000 k-points is used in the  irreducible BZ for the calculations of energy band structures.
\begin{figure}
  \includegraphics[width=7cm]{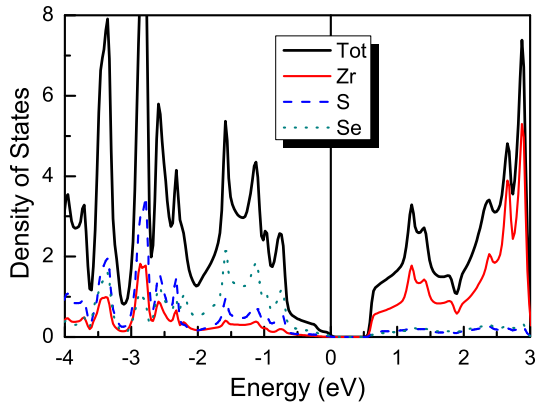}
  \caption{(Color online) The total and atomic partial  DOS of  ZrSSe monolayer using GGA+SOC.}\label{dos}
\end{figure}

The  lattice thermal conductivity is performed
by using Phono3py+VASP codes\cite{pv1,pv2,pv3,pv4}.
With the plane-wave-cut-off energy of 500 eV, the GGA-PBE\cite{pbe} is used  for the exchange-correlation functional.
The energy  convergence criteria is   $10^{-8}$ eV.
The  lattice thermal conductivity is calculated  by solving linearized phonon Boltzmann equation with single-mode relaxation time approximation  (RTA),  as implemented in the Phono3py code\cite{pv4}. The lattice thermal conductivity can be attained by the following formula:
\begin{equation}\label{eq0}
    \kappa=\frac{1}{NV_0}\sum_\lambda \kappa_\lambda=\frac{1}{NV_0}\sum_\lambda C_\lambda \nu_\lambda \otimes \nu_\lambda \tau_\lambda
\end{equation}
where $\lambda$ is phonon mode, $N$ is the total number of q points sampling BZ, $V_0$ is the volume of a unit cell, and  $C_\lambda$,  $ \nu_\lambda$, $\tau_\lambda$   is the specific heat,  phonon velocity,  phonon lifetime.
The phonon lifetime $\tau_\lambda$ can be calculated by  phonon linewidth $2\Gamma_\lambda(\omega_\lambda)$ of the phonon mode
$\lambda$:
\begin{equation}\label{eq0}
    \tau_\lambda=\frac{1}{2\Gamma_\lambda(\omega_\lambda)}
\end{equation}
The $\Gamma_\lambda(\omega)$  takes the form analogous to the Fermi golden rule:
\begin{equation}
\begin{split}
   \Gamma_\lambda(\omega)=\frac{18\pi}{\hbar^2}\sum_{\lambda^{'}\lambda^{''}}|\Phi_{-\lambda\lambda^{'}\lambda^{''}}|^2
   [(f_\lambda^{'}+f_\lambda^{''}+1)\delta(\omega
    -\omega_\lambda^{'}-\\\omega_\lambda^{''})
   +(f_\lambda^{'}-f_\lambda^{''})[\delta(\omega
    +\omega_\lambda^{'}-\omega_\lambda^{''})-\delta(\omega
    -\omega_\lambda^{'}+\omega_\lambda^{''})]]
\end{split}
\end{equation}
in which $f_\lambda$ is the phonon equilibrium occupancy and
$\Phi_{-\lambda\lambda^{'}\lambda^{''}}$
is the strength of interaction among the three phonons $\lambda$, $\lambda^{'}$,
and $\lambda^{''}$ involved in the scattering.
Based on the supercell approach  with finite atomic displacement
of 0.03 $\mathrm{{\AA}}$,  the second (third)-order interatomic force constants (IFCs) can be attained
by using   a 5 $\times$ 5 $\times$ 1 (3 $\times$ 3 $\times$ 1)  supercell   with k-point meshes of 2 $\times$ 2 $\times$ 1 (3 $\times$ 3 $\times$ 1).
According to the harmonic IFCs, phonon dispersions can be attained by  Phonopy package\cite{pv5}.
 To compute lattice thermal conductivities, the reciprocal spaces of the primitive cells  are sampled using the 100 $\times$ 100 $\times$ 1 meshes.

It is noted that, for 2D material, the calculated  electrical and thermal conductivities  depend on the length of unit cell along z direction\cite{2dl}.  They  should be normalized by multiplying $Lz/d$, in which  $Lz$ is the length of unit cell along z direction,  and d is the thickness of 2D material. It is well known that the d  is not well defined like graphene.  In this work, the $Lz$=20 $\mathrm{{\AA}}$  is used as $d$. By $\kappa$ $\times$ $d$,  the thermal sheet conductance can be attained.

\section{MAIN CALCULATED RESULTS AND ANALYSIS}
The structure of Janus  monolayer ZrSSe  is similar to   monolayer  $\mathrm{ZrS_2}$ with the 1T phase,
which contains three atomic sublayers with Zr layer sandwiched between S and Se layers.
The organized Janus monolayer ZrSSe can be attained  by
fully replacing one of two  S  layers with Se  atoms in  $\mathrm{ZrS_2}$ monolayer.
Compared with $\mathrm{ZrS_2}$ monolayer, the Janus
monolayer ZrSSe lacks the reflection symmetry with respect to the central metal Zr atoms.
 The schematic crystal structure is given  in \autoref{st}, which belongs to the space group of $P3m1$ (No.156), being lower than one of
  $\mathrm{ZrS_2}$  monolayer ($P\bar{3}m1$[No.164]). To avoid spurious interaction, the unit cell  is built with the vacuum region of larger than 15 $\mathrm{{\AA}}$. The optimized lattice constants within GGA-PBE are $a$=$b$=3.74 $\mathrm{{\AA}}$, which are
about 1.6\%  higher than those of  $\mathrm{ZrS_2}$ monolayer ($a$=$b$=3.68 $\mathrm{{\AA}}$). It is found that the bond length of Zr-S between ZrSSe and $\mathrm{ZrS_2}$ monolayers is almost the same, but the bond length of Zr-Se in ZrSSe monolayer is longer than that of Zr-S. The related data are listed in \autoref{tab}.
\begin{figure*}
  \includegraphics[width=12cm]{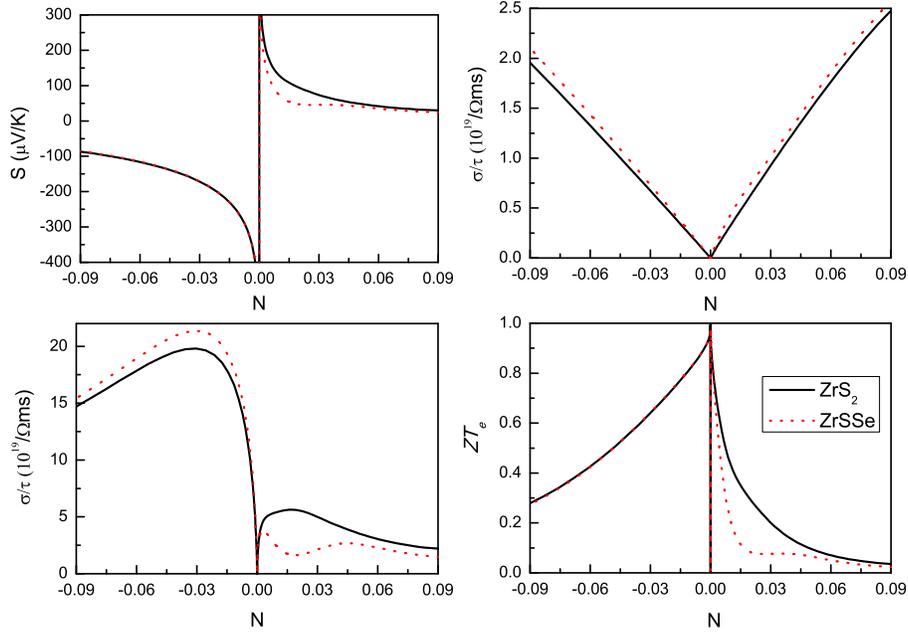}
  \caption{(Color online) At room temperature,   Seebeck coefficient S, electrical conductivity with respect to scattering time  $\mathrm{\sigma/\tau}$,    power factor with respect to scattering time $\mathrm{S^2\sigma/\tau}$ and $ZT_e$ of  $\mathrm{ZrS_2}$ and ZrSSe as a function of doping level (N)  using  GGA+SOC.}\label{s}
\end{figure*}
\begin{figure}
    \includegraphics[width=7cm]{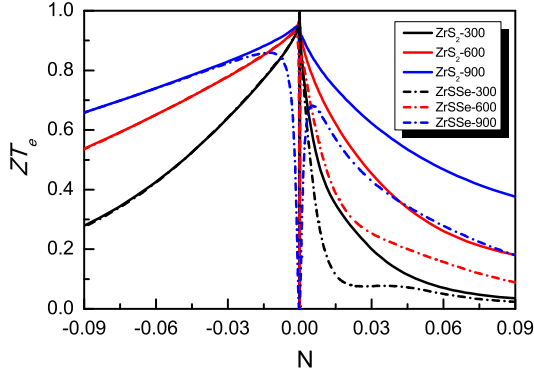}
  \caption{(Color online)At 300, 600 and 900 K, the  $ZT_e$ of  $\mathrm{ZrS_2}$ and ZrSSe monolayers as a function of doping level (N)  using  GGA+SOC.}\label{s-1}
\end{figure}

To confirm the dynamical stability, phonon dispersion of
Janus  monolayer ZrSSe is plotted in \autoref{phband}. It is clearly seen that no
imaginary vibrational frequency is observed in the first BZ, which suggests that monolayer ZrSSe is dynamically stable.
The mechanical stability of monolayer ZrSSe can be confirmed by
calculating the  elastic constants. Due to the hexagonal symmetry, there are two independent elastic
constants $C_{11}$=$C_{22}$ and $C_{12}$, and the $C_{66}$=($C_{11}$-$C_{12}$)/2. The elastic
constants $C_{ij}$ of $\mathrm{ZrS_2}$ and ZrSSe monolayers are listed in \autoref{tab1}, which satisfy the  Born  criteria of mechanical stability, namely
\begin{equation}\label{e1}
C_{11}>0,~~ C_{66}>0
\end{equation}
The 2D Young¡¯s moduli  in the
Cartesian [10] and [01] directions are given\cite{ela}
\begin{equation}\label{e1}
Y^{2D}_{[10]}=\frac{C_{11}C_{22}-C_{12}^2}{C_{22}},~~ Y^{2D}_{[01]}=\frac{C_{11}C_{22}-C_{12}^2}{C_{11}}
\end{equation}
The corresponding Poisson's ratios can be attained by
\begin{equation}\label{e1}
\nu^{2D}_{[10]}=\frac{C_{12}}{C_{22}},~~ \nu^{2D}_{[01]}=\frac{C_{12}}{C_{11}}
\end{equation}
and the 2D shear modulus is
\begin{equation}\label{e1}
G^{2D}=C_{66}
\end{equation}
According to \autoref{tab1}, the elastic constants, shear modulus, Young's modulus of ZrSSe monolayer  are slightly smaller than ones of $\mathrm{ZrS_2}$ monolayer. This indicates that ZrSSe monolayer is more flexible  than   $\mathrm{ZrS_2}$ monolayer due to  smaller Young's
modulus. The smaller Young's modulus may be resulting from the weaker Zr-Se bond strength.
The calculated Young's modulus and  Poisson's ratio of  $\mathrm{ZrS_2}$ monolayer agree well with previous theoretical values\cite{rsc}, which  ensures the reliability of our results. The elastic constants, shear modulus and Young's modulus of $\mathrm{MoS_2}$ and MoSSe monolayers\cite{n1} are also listed in \autoref{tab1}. Similar to $\mathrm{ZrS_2}$ and ZrSSe monolayers, the related elastic quantities of MoSSe monolayer are smaller than ones of  $\mathrm{MoS_2}$ monolayer. Compared to  MoSSe, the  in-plane strain engineering of  large magnitude in ZrSSe monolayer can be easily achieved due to small Young's modulus, which is very important for tuning physical properties of ZrSSe monolayer by strain.
\begin{figure}
  \includegraphics[width=8.0cm]{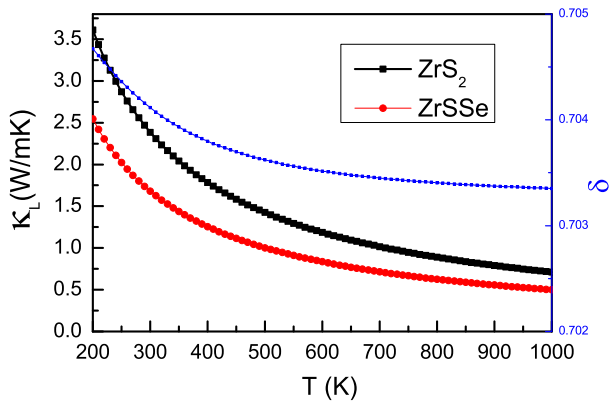}
  \caption{(Color online) the lattice thermal conductivities of  $\mathrm{ZrS_2}$ and ZrSSe monolayers  as a function of temperature using GGA-PBE, and $\kappa_L$(ZrSSe)/$\kappa_L$($\mathrm{ZrS_2}$). }\label{mkl}
\end{figure}
\begin{figure}
  \includegraphics[width=8.0cm]{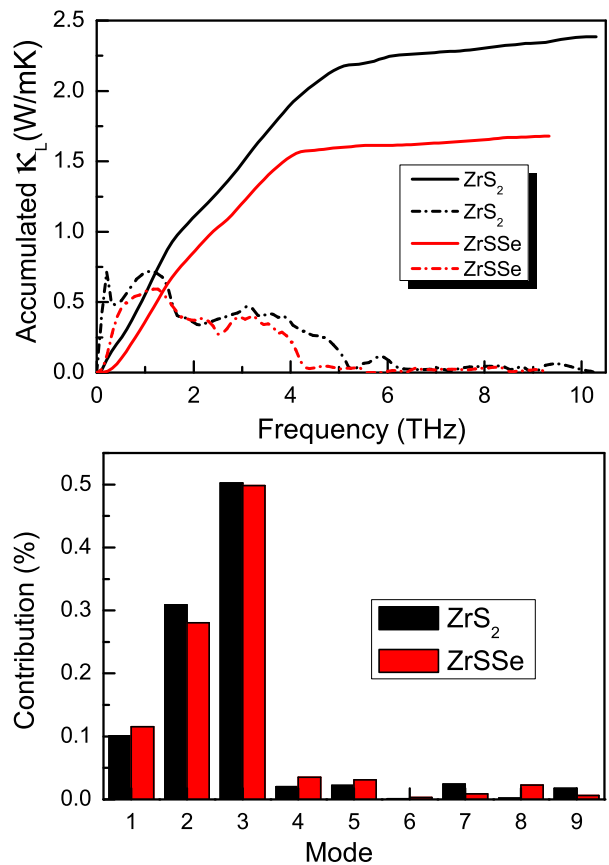}
  \caption{(Color online)Top: the accumulated lattice thermal conductivities of  $\mathrm{ZrS_2}$ and ZrSSe monolayers, and the derivatives. Bottom: phonon modes contributions toward total lattice thermal conductivity (300 K). 1 represents ZA branch, 2 for TA branch, 3 for LA branch and 4, 5, 6, 7, 8, 9 for optical branches.}\label{mode}
\end{figure}

The SOC has important effects on electronic structures and electronic   transport coefficients in semiconducting TMD monolayers\cite{q14,qgsd}.
Here, GGA+SOC is employed to study energy band structures  and  electronic transport properties of  $\mathrm{ZrS_2}$ and ZrSSe monolayers, and their energy band structures are plotted in \autoref{band}. It is found that both  $\mathrm{ZrS_2}$ and ZrSSe  monolayers are indirect band gap semiconductors with the valence band maximum (VBM) at the $\Gamma$ point and  conduction band minimum (CBM)  at the M point. The band gap (0.60 eV) of ZrSSe monolayer is smaller than that (1.12 eV) of $\mathrm{ZrS_2}$ monolayer, but  the spin-orbit splitting at  the ${\Gamma}$ point near the Fermi level in the valence bands is larger for ZrSSe monolayer (0.21 eV) than $\mathrm{ZrS_2}$  monolayer (0.09 eV).
Due to both inversion and time-reversal symmetries, all the bands of $\mathrm{ZrS_2}$ monolayer are doubly degenerate. However, It is clearly seen that double degenerate bands of ZrSSe monolayer  are removed due to lack of inversion  symmetry.  From density of states (DOS) of ZrSSe  monolayer in \autoref{dos}, the valence bands near the fermi level are composed of S-p  and Se-p states, while Zr-d states have main  contributions in the conduction bands.
\begin{figure*}
  \includegraphics[width=14.0cm]{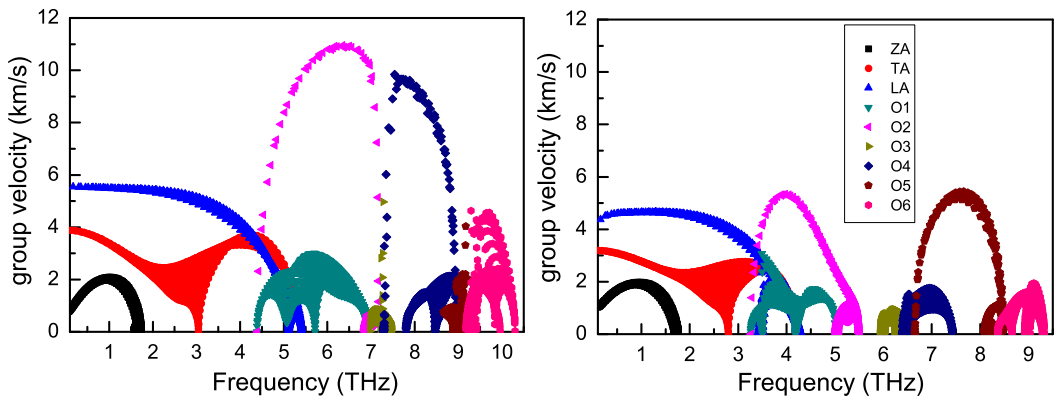}
  \caption{(Color online) the phonon mode group velocities of  $\mathrm{ZrS_2}$ and ZrSSe  monolayers in the first BZ. }\label{v}
\end{figure*}

After the energy band structures are completed, the  electronic   transport coefficients of  $\mathrm{ZrS_2}$ and ZrSSe monolayers can be attained  using CSTA Boltzmann theory.  The Seebeck coefficient S,  electrical conductivity with respect to scattering time  $\mathrm{\sigma/\tau}$,  power factor with respect to scattering time $\mathrm{S^2\sigma/\tau}$  and $ZT_e$ as a function of doping level using  GGA+SOC at room temperature are plotted in \autoref{s}. The electronic thermal conductivity  $\mathrm{\kappa_e}$ is connected to the electrical
conductivity  $\mathrm{\sigma}$ by the Wiedemann-Franz law:
\begin{equation}\label{eq2}
  \kappa_e=L\sigma T
\end{equation}
where $L$ is the Lorenz number. It is noted that the S is independent of $\tau$, while electrical conductivity  and power factor depend  on  $\tau$. The  $ZT_e=S^2\sigma T/\kappa_e$ as an upper limit of $ZT$ is also independent of $\tau$,  taking no account of $\kappa_L$. Within the framework of  rigid band approach, the n- or p-type doping  can be simulated by simply shifting  Fermi level into conduction  or valence bands.  It is found that n-type S between  $\mathrm{ZrS_2}$ and ZrSSe monolayers are pretty much the same,  which is due to similar profile of conduction bands between  $\mathrm{ZrS_2}$ and ZrSSe monolayers. However, the p-type S of ZrSSe monolayer is lower than that of $\mathrm{ZrS_2}$ monolayer in low doping level, which can be explained by enhanced  spin-orbit splitting in the
valence bands for ZrSSe monolayer,   reducing  the strength of orbital  degeneracies. The  $\mathrm{\sigma/\tau}$ of ZrSSe monolayer is slightly higher than that of   $\mathrm{ZrS_2}$ monolayer in considered doping level range. For  $\mathrm{S^2\sigma/\tau}$,  n-type doping of ZrSSe monolayer is higher than that of $\mathrm{ZrS_2}$ monolayer, while p-type doping is opposite.  The relation of $ZT_e$ between $\mathrm{ZrS_2}$ and ZrSSe monolayers is the same with that of S.
The temperature dependence of $ZT_e$ of  $\mathrm{ZrS_2}$ and ZrSSe monolayers is also plotted in \autoref{s-1}. It is clearly seen that the n-type $ZT_e$  between $\mathrm{ZrS_2}$ and ZrSSe monolayers is almost the same in considered doping and temperature range  except low doping at 900 K.
If  $\kappa_L$ of ZrSSe monolayer  was lower than that of $\mathrm{ZrS_2}$ monolayer, the ZrSSe monolayer would have better n-type  thermoelectric properties than $\mathrm{ZrS_2}$.

Next, we investigate the phonon transports of  $\mathrm{ZrS_2}$ and ZrSSe monolayers. Firstly, the phonon dispersions of $\mathrm{ZrS_2}$ are also plotted in \autoref{phband}, together with atomic  partial DOS, which are consistent with available theoretical results\cite{n4}.
The phonon dispersions determine   the allowed three-phonon scattering processes.
 Due to three atoms in the unit cell of $\mathrm{ZrS_2}$ and ZrSSe monolayers, there are 3 acoustic and 6 optical phonon branches.
The longitudinal acoustic (LA) and transverse acoustic (TA) branches  are linear near the $\Gamma$ point, while the z-direction acoustic (ZA) branch is quadratic, which shares general features of 2D materials\cite{q12,q15,q16,q16-1,q21,q22,q14,q14-1,q27-1,q27-2,q27-3}.  Compared to $\mathrm{ZrS_2}$ monolayer,  the phonon dispersions of
TA and LA modes of ZrSSe monolayer become softened, while ZA mode has little change. From $\mathrm{ZrS_2}$ to ZrSSe monolayer,  the whole optical branches   move toward lower energy. These mean that ZrSSe monolayer has smaller  group velocities than $\mathrm{ZrS_2}$ monolayer. Unlike $\mathrm{MoS_2}$ monolayer\cite{new2}, no  acoustic-optical (a-o) gap  is observed for both $\mathrm{ZrS_2}$ and ZrSSe monolayers, which produce important effects on acoustic+acoustic$\rightarrow$optical (aao) scattering. For  $\mathrm{ZrS_2}$ monolayer, the  optical modes are mainly from S vibrations, while the acoustic branches are due to the vibrations of Zr.   For ZrSSe monolayer, the high-frequency optical modes are mainly due to S vibrations, while the low-frequency optical and acoustic branches are due to the vibrations of Zr and Se atoms.
\begin{figure*}
  \includegraphics[width=14.0cm]{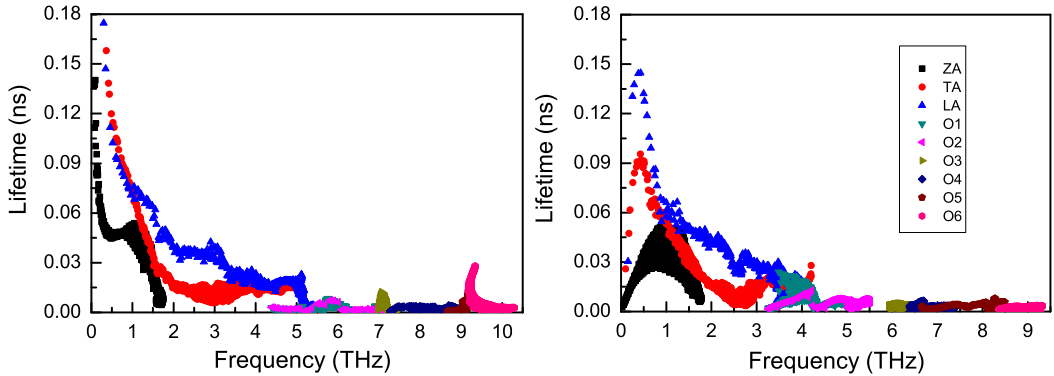}
  \caption{(Color online)the phonon mode phonon lifetimes (300 K)  of  $\mathrm{ZrS_2}$ and ZrSSe monolayers in the first BZ. }\label{t}
\end{figure*}
\begin{figure}
  \includegraphics[width=8.0cm]{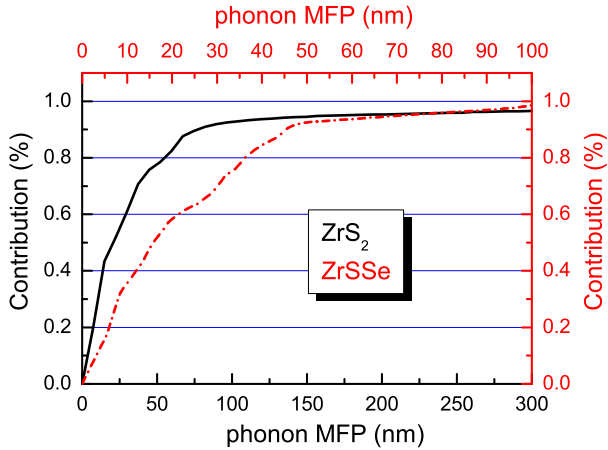}
  \caption{(Color online) the cumulative lattice thermal conductivity of $\mathrm{ZrS_2}$ and ZrSSe monolayers   divided by their total lattice thermal conductivity with respect to phonon MFP at room temperature.}\label{mfp}
\end{figure}

Based on  harmonic and anharmonic IFCs, the intrinsic lattice thermal conductivities  of $\mathrm{ZrS_2}$ and ZrSSe monolayers as a function of temperature  are plotted in \autoref{mkl}, along with $\kappa_L$(ZrSSe)/$\kappa_L$($\mathrm{ZrS_2}$). For both cases, the $\kappa_L$ follows a $T^{-1}$ with the increasing temperature, which is due to enhanced phonon-phonon scattering. At 300 K, the lattice  thermal conductivity  is 2.39 $\mathrm{W m^{-1} K^{-1}}$ for $\mathrm{ZrS_2}$ monolayer and 1.68  $\mathrm{W m^{-1} K^{-1}}$ for ZrSSe monolayer with the same thickness 20 $\mathrm{{\AA}}$.
The thermal sheet conductance can be used to make fair comparison for thermal transport properties of 2D materials\cite{2dl}, and  the corresponding  thermal sheet conductance is  47.8 $\mathrm{W K^{-1}}$ for $\mathrm{ZrS_2}$ monolayer and 33.6 $\mathrm{W K^{-1}}$ for ZrSSe monolayer. In the temperature range investigated, the thermal conductivities of ZrSSe monolayer are about 70\% of ones of  $\mathrm{ZrS_2}$ monolayer. Combining $ZT_e$ with $\kappa_L$, ZrSSe monolayer may have better thermoelectric properties than $\mathrm{ZrS_2}$ monolayer in n-type doping. For both $\mathrm{ZrS_2}$ and ZrSSe monolayers,  the accumulated lattice thermal conductivities,  the derivatives and the contribution of each phonon
mode to the total lattice thermal conductivity are shown in \autoref{mode}. It is clearly seen that  the three acoustic phonon branches contribute
mostly to the lattice  thermal conductivity, around 91.26\% for  $\mathrm{ZrS_2}$ monolayer and 89.38\% for ZrSSe monolayer. It is found that the  contribution gradually increases from ZA to TA to LA branch  in acoustic branches, about 10\% for ZA branch, 30\% for TA branch and 50\% for LA branch.

To investigate the origin of lower $\kappa_L$ for ZrSSe monolayer than  $\mathrm{ZrS_2}$ monolayer, the mode level phonon group velocities
and lifetimes are plotted in \autoref{v} and \autoref{t}.  The group velocities can be calculated by phonon dispersions from second-order harmonic IFCs\cite{pv4}. The three-phonon scattering rate can be calculated by third-order anharmonic IFCs, and then  the phonon lifetimes can be attained\cite{pv4}.
Due to  quadratic dispersion of ZA branch near the $\Gamma$ point,  the  group velocities  of  ZA branch are smaller  than  ones of LA and TA branches.  The largest  group velocity of $\mathrm{ZrS_2}$ monolayer near $\Gamma$ point is
 0.84  $\mathrm{km s^{-1}}$ for ZA, 3.89  $\mathrm{km s^{-1}}$ for LA,  and 5.54  $\mathrm{km s^{-1}}$ for TA.
 For ZrSSe monolayer,  the corresponding value is 1.06 $\mathrm{km s^{-1}}$, 3.20 $\mathrm{km s^{-1}}$ and 4.37 $\mathrm{km s^{-1}}$.
 It is found that  group velocities of ZA branch between $\mathrm{ZrS_2}$ and ZrSSe  monolayers are almost the same.  For other branches, it is clearly seen that  most of group velocities are lower for ZrSSe  monolayer than $\mathrm{ZrS_2}$ monolayer, which can give rise to decreasing  lattice thermal conductivity.
 These can be explained by  phonon softening, which leads to lower  group velocity for ZrSSe monolayer than $\mathrm{ZrS_2}$ monolayer.
 On the other hand, the most phonon lifetimes of  ZrSSe monolayer for ZA branch are shorter  than ones of  $\mathrm{ZrS_2}$ monolayer, which will
benefit the  low lattice thermal conductivity. The separation between ZA acoustic branch and  optical
phonon branches is smaller for ZrSSe monolayer than $\mathrm{ZrS_2}$ monolayer, which leads to much more frequent
scattering between ZA acoustic modes and optical modes, resulting in  shorter phonon lifetimes.  Compared to  $\mathrm{ZrS_2}$ monolayer, the lower $\kappa_L$ of ZrSSe monolayer is mainly due to smaller group  velocities and shorter phonon lifetimes of ZA mode.

The cumulative lattice thermal conductivity   with respect to phonon mean free
paths (MFP) can provide information about
the contributions of phonons with different MFP to  the total thermal conductivity.
At room temperature,  the cumulative lattice thermal conductivity divided by total lattice thermal conductivity  of $\mathrm{ZrS_2}$ and ZrSSe monolayers with respect to phonon MFP are plotted in  \autoref{mfp}.
For $\mathrm{ZrS_2}$ (ZrSSe) monolayer, phonons with MFP smaller than 55 (37)
nm contribute  around 80\% to total  lattice thermal conductivity. This is consistent with lower lattice thermal conductivity for ZrSSe monolayer than $\mathrm{ZrS_2}$ monolayer. The stronger intrinsic phonon scattering for ZrSSe monolayer than $\mathrm{ZrS_2}$ monolayer can induce lower lattice thermal conductivity by causing  phonons to have shorter MFP.

\section{Discussions and Conclusion}
\begin{figure}
  \includegraphics[width=8cm]{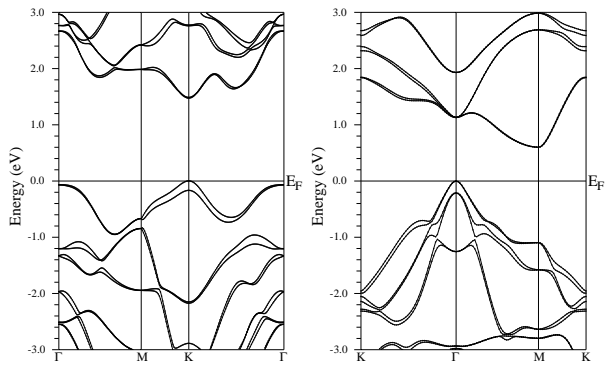}
  \caption{The energy band structures of  MoSSe (Left) and ZrSSe (Right) monolayers  using GGA+SOC.}\label{band1}
\end{figure}

Recently, the sandwiched S-Mo-Se structure (Janus MoSSe monolayer) has been synthesized  by fully replacing the top  Se layers with S atoms within $\mathrm{MoSe_2}$ monolayer\cite{q7-1}.
The MoSSe and ZrSSe monolayers  have the same space group of $P3m1$ (No.156), but the different stacking of
S and Se sublayers leads to two crystal structures, namely 1T structure (ZrSSe) and 2H structure (MoSSe),  which is due to different ionicity.
From \autoref{band1}, the  indirect gap of ZrSSe monolayer (0.60 eV) is very smaller than direct gap of MoSSe monolayer(1.47 eV).  However,  near the Fermi level, the spin-orbit splitting of ZrSSe monolayer (0.21 eV) at  the ${\Gamma}$ point in the valence bands is larger  than that of  MoSSe monolayer (0.17 eV) at the K point. The predicted elastic stiffness coefficient of Janus MoSSe monolayer is  126.8 N/m for $C_{11}$,  27.4 N/m for $C_{12}$\cite{n1}, which are larger than ones of ZrSSe monolayer ($C_{11}$=68.8 N/m, $C_{12}$=14.7 N/m). Compared to MoSSe monolayer, ZrSSe monolayer is more flexible with smaller elastic stiffness coefficients, and the increased flexible nature  makes ZrSSe monolayer to be a good choice for large magnitude  strain control.

Strain is a very effective method to tune electronic structures, topological  and transport  properties of 2D materials\cite{q27-1,q27-2,q27-3,n4,t9}. Indirect-to-direct band gap transition of $\mathrm{ZrS_2}$ monolayer by strain  has been predicted by first-principles calculations\cite{rsc}, and the
band structure can be remarkably modified.  For  $\mathrm{ZrS_2}$ monolayer, the greatly enhanced thermoelectric performance caused  by the biaxial tensile strain  has been predicted by first-principles calculations combined with the
Boltzmann transport theory\cite{n4}.     It is found that  tensile strain can  increase  the Seebeck coefficient of $\mathrm{ZrS_2}$ monolayer by  bands converge, and
decrease the lattice  thermal conductivity by  reducing group velocities of the TA and LA modes. For bulk  $\mathrm{ZrS_2}$, the enhanced thermoelectric performance, caused by strain, has also been predicted by ab initio calculations and semiclassical Boltzmann transport theory, which is due to convergence of separate orbits\cite{new3}. The ZrSSe and $\mathrm{ZrS_2}$ monolayers have similar elastic stiffness coefficients, electronic structures and phonon behaviour, so it is possible to improve thermoelectric performance of ZrSSe monolayer  by strain engineering.

In summary, using first-principles calculations and semiclassical Boltzmann transport theory, we
investigate  the stability, mechanical, electronic and transport properties  of ZrSSe monolayer.
It is proved that ZrSSe monolayer is dynamically  and  mechanically  stable by  phonon dispersion and  Born  criteria of mechanical stability.
The ZrSSe monolayer is an indirect gap semiconductor with remarkable spin-orbit splitting.
It is found that n-type $ZT_e$ between ZrSSe and $\mathrm{ZrS_2}$ monolayers is almost the same due to identical Seebeck coefficient.
However, the $\kappa_L$ of ZrSSe monolayer is lower than that of $\mathrm{ZrS_2}$ monolayer, which is due to lower  group velocities and  shorter phonon lifetimes of ZA phonon mode for ZrSSe monolayer than $\mathrm{ZrS_2}$ monolayer.  Therefore, the ZrSSe monolayer may have better n-type thermoelectric performance than $\mathrm{ZrS_2}$ monolayer.
The n-type  doping of the
ZrSSe monolayer can be realized by adsorption of small molecules,  substituting site atoms and electrolyte gating\cite{new4,new5,new6}.
The ZrSSe monolayer is a potential candidate for thermoelectric application, and our works can stimulate further experimental works to synthesize  ZrSSe monolayer.

\begin{acknowledgments}
This work is supported by the National Natural Science Foundation of China (Grant No.11404391).  We are grateful to the Advanced Analysis and Computation Center of CUMT for the award of CPU hours to accomplish this work.
\end{acknowledgments}


\begin{references}
\bibitem{q0}T. M. Tritt and M. A. Subramanian, MRS Bulletin \textbf{31},  188 (2006).
\bibitem{q1}G. J. Snyder and  E. S. Toberer, Nature Materials \textbf{7}, 105  (2008).

\bibitem{q4}A. J. Minnich,   M. S. Dresselhaus,   Z. F. Ren and   G. Chen, Energy Environ. Sci. \textbf{2}, 466 (2009).


\bibitem{q6}T. C. Harman, P. J. Taylor, M. P. Walsh and  B. E. LaForge, Science  \textbf{297},  2229 (2002).

\bibitem{new1}M. Cutler, J. F. Leavy, and  R. L.  Fitzpatrick, Phys. Rev.  \textbf{133}, A1143 (1964).


\bibitem{q2}L. D. Hicks and M. S. Dresselhaus, Phys. Rev. B \textbf{47}, 12727  (1993).

\bibitem{q3}L. D. Hicks and M. S. Dresselhaus, Phys. Rev. B \textbf{47}, 16631(R) (1993).





\bibitem{q7}M. Chhowalla,	H. S. Shin,	G. Eda,	L. J.  Li,	K. P.  Loh	and H. Zhang, Nature Chemistry \textbf{5}, 263  (2013).

\bibitem{q7-1}A. Y. Lu, H. Y. Zhu, J. Xiao et al., Nature Nanotechnology \textbf{12}, 744 (2017).

\bibitem{q8}R. X.  Fei, W. B. Li, J. Li and L. Yang, Appl. Phys. Lett. \textbf{107}, 173104 (2015).

\bibitem{q9}S. L. Zhang  et al.,  Angew. Chem. \textbf{128}, 1698 (2016).


\bibitem{q10}J. P. Ji  et al.,  Nat. Commun. \textbf{7}, 13352 (2016).


\bibitem{q11}S. Balendhran, S. Walia, H. Nili, S. Sriram and M.Bhaskaran, small \textbf{11},  640 (2015).


\bibitem{q12}W. Huang, H. X. Da and G. C.  Liang,  J. Appl. Phys.  \textbf{113}, 104304 (2013).




\bibitem{q15}G. Qin, Z. Qin, W. Fang, L. Zhang, S. Yue, Q. Yan, M. Hu and G. Su, Nanoscale \textbf{8}, 11306 (2016).

\bibitem{q16}S. D. Guo and Y. H. Wang, J. Appl. Phys. \textbf{121}, 034302 (2017).

\bibitem{q16-1}D. C. Zhang, A. X. Zhang,  S. D. Guo and Y. F. Duan, RSC Adv.   \textbf{7}, 24537 (2017).

\bibitem{q21}L. M.  Sandonas,D. Teich, R. Gutierrez, T. Lorenz, A. Pecchia, G. Seifert  and G. Cuniberti, J. Phys. Chem. C  \textbf{120}, 18841 (2016).

\bibitem{q22}L. Cheng, H. J. Liu, X. J. Tan, J. Zhang, J. Wei, H. Y.  Lv, J.  Shi and X. F. Tang, J. Phys. Chem. C \textbf{118}, 904 (2014).

\bibitem{q14}S. D. Guo, J. Mater. Chem. C  \textbf{4}, 9366 (2016).

\bibitem{q14-1}S. D.  Guo, A. X. Zhang and H. C. Li, Nanotechnology \textbf{28},  445702 (2017).

\bibitem{q14-2}G. P. Li, K. L. Yao and G. Y. Gao, Nanotechnology \textbf{29}, 015204 (2018).

\bibitem{q14-3}G. P. Li, G. Q. Ding and G. Y. Gao, J. Phys.: Condens. Matter \textbf{29}, 015001 (2017).

\bibitem{q13}J.  Wu  et al.  Nano Lett. \textbf{14}, 2730  (2014).


\bibitem{q23}B.  Peng, H.  Zhang, H. Z. Shao, Y. F.  Xu, G.  Ni, R. J.  Zhang  and H. Y.  Zhu, Phys. Rev. B  \textbf{94}, 245420 (2016).


\bibitem{q24}H. M. Wang,  G. Z. Qin,  G. J. Li,  Q.  Wang  and  M. Hu, Phys. Chem. Chem. Phys. \textbf{19}, 12882 (2017).


\bibitem{q25} Z. Z. Qin, G. Z. Qin, X. Zuo, Z. H. Xiong and M. Hu, Nanoscale  \textbf{9}, 4295 (2017).


\bibitem{q26}S. D. Guo and J. T. Liu,  Phys. Chem. Chem. Phys. \textbf{19}, 31982 (2017).



\bibitem{q27-1}Y. D. Kuang, L. Lindsay, S. Q. Shic and G. P. Zheng, Nanoscale  \textbf{8}, 3760 (2016).



\bibitem{q27-2}H. K. Liu,  G. Z. Qin, Y. Lin and M. Hu,  Nano Lett.  \textbf{16}, 3831 (2016).

\bibitem{q27-3}A. X. Zhang, J. T. Liu, S. D. Guo  and H. C. Li, Phys. Chem. Chem. Phys. \textbf{19}, 14520  (2017).

\bibitem{n1}L. Dong, J.  Lou and V.  B. Shenoy, ACS Nano \textbf{11}, 8242 (2017).

\bibitem{n2}F. P.  Li, W. Wei, P. Zhao, B. B. Huang and Y.  Dai, J. Phys. Chem. Lett.   \textbf{8},  5959 (2017).

\bibitem{n3}Z. Zeng, Z. Yin, X. Huang, H. Li, Q. He, G. Lu, F. Boey and
H. Zhang, Angew. Chem., Int. Ed. \textbf{50}, 11093 (2011).


\bibitem{n4}H. Y. Lv,   W. J. Lu,   D. F. Shao,  H. Y. Lub and   Y. P. Sun, J. Mater. Chem. C \textbf{4}, 4538 (2016).

\bibitem{1}P. Hohenberg and W. Kohn, Phys. Rev. \textbf{136},
B864 (1964); W. Kohn and L. J. Sham, Phys. Rev. \textbf{140},
A1133 (1965).

\bibitem{2}P. Blaha, K. Schwarz, G. K. H. Madsen, D. Kvasnicka
 and J. Luitz, WIEN2k, an Augmented Plane Wave
+ Local Orbitals Program for Calculating Crystal Properties
(Karlheinz Schwarz Technische Universit\"at Wien, Austria) 2001,
ISBN 3-9501031-1-2



\bibitem{pbe}J. P. Perdew, K. Burke and M. Ernzerhof, Phys. Rev. Lett. \textbf{77}, 3865 (1996).

\bibitem{10}A. H. MacDonald, W. E. Pickett and D. D. Koelling, J. Phys. C \textbf{13}, 2675 (1980).

\bibitem{11}D. J. Singh and L. Nordstrom, Plane Waves, Pseudopotentials and the LAPW
Method, 2nd Edition (Springer, New York, 2006).

\bibitem{12}J. Kunes, P. Novak, R. Schmid, P. Blaha and
K. Schwarz, Phys. Rev. B \textbf{64}, 153102 (2001).

\bibitem{so}D. D. Koelling, B. N. Harmon, J. Phys. C Solid State Phys.  \textbf{10}, 3107 (1977).



\bibitem{b}G. K. H. Madsen and D. J. Singh, Comput. Phys. Commun. \textbf{175}, 67
(2006).



\bibitem{pv1} G. Kresse, J. Non-Cryst. Solids \textbf{193}, 222 (1995).

\bibitem{pv2} G. Kresse and J. Furthm$\ddot{u}$ller, Comput. Mater. Sci. 6, \textbf{15} (1996).

\bibitem{pv3} G. Kresse and D. Joubert, Phys. Rev. B \textbf{59}, 1758 (1999).

\bibitem{pv4}A. Togo, L. Chaput and I. Tanaka, Phys. Rev. B \textbf{91}, 094306 (2015).

\bibitem{pv5}A. Togo, F. Oba, and I. Tanaka, Phys. Rev. B \textbf{78}, 134106
(2008).

\bibitem{2dl}X. F. Wu, V. Varshney et al., Chem. Phys. Lett. \textbf{669}, 233 (2017).

\bibitem{ela}R. C. Andrew, R. E. Mapasha, A. M. Ukpong and N. Chetty, Phys. Rev. B \textbf{85}, 125428 (2012).


\bibitem{rsc}Y. Li, J. Kang and J. B. Li, RSC Adv. \textbf{4}, 7396 (2014).

\bibitem{qgsd}S. D. Guo and J. L. Wang, Semicond. Sci. Tech. \textbf{31}, 095011 (2016).

\bibitem{new2}X. K. Gu and R. G. Yang,  Appl. Phys. Lett. \textbf{105}, 131903 (2014).

\bibitem{t9}S. L.  Zhang, M. Q. Xie, B. Cai, H. J. Zhang, Y. D. Ma, Z. F. Chen, Z. Zhu,
Z. Y. Hu, and H. B. Zeng, Phys. Rev. B \textbf{93}, 245303 (2016).


\bibitem{new3}G. Q. Ding, J. F. Chen, K. L. Yao and G. Y. Gao, New J. Phys.  \textbf{19}, 073036 (2017).



\bibitem{new4}D. Kiriya, M. Tosun, P. Zhao, J. S. Kang and A. Javey, J. Am.
Chem. Soc. \textbf{136}, 7853 (2014).

\bibitem{new5}M. R. Laskar, D. N. Nath, L. Ma et al., Appl. Phys. Lett. \textbf{104}, 092104 (2014).

\bibitem{new6}J. T. Ye, Y. J. Zhang, R. Akashi, M. S. Bahramy, R. Arita and
Y. Iwasa, Science \textbf{338}, 1193 (2012).

\end{references}
\end{document}